\documentclass{emulateapj}
\usepackage{apjfonts}
\usepackage{times}
\usepackage{epsfig}
\usepackage{color}
\usepackage{tikz}
\usepackage{comment}

\def\aap{ A\&A}

\def\apjs{ApJ Supp}
\def\apjl{ ApJL}

\shorttitle{BH-XRB: Formation Puzzle}
\shortauthors{Wiktorowicz, Belczynski \& Maccarone}

\newcommand{\msun}{\ensuremath{\,\mathrm{M}_\odot}}

\newcommand{\myr}{\ensuremath{\,\mathrm{Myr}}}
\newcommand{\gyr}{\ensuremath{\,\mathrm{Gyr}}}
\newcommand{\rsun}{\ensuremath{\,\mathrm{R}_\odot}}
\newcommand{\startrack}{{\tt StarTrack }}
\newcommand{\s}{\ensuremath{\S\,}}
\newcommand{\msy}{\ensuremath{\msun\mathrm{\; yr}^{-1}}}
 
\begin{document}

\title{Black Hole X-ray Transients: the Formation Puzzle}
\author{G.\ Wiktorowicz$^1$, K.\ Belczynski$^{1,2}$, T.J.\ Maccarone$^3$}
\altaffiltext{1}{Astronomical Observatory,University of Warsaw, 
                 Al. Ujazdowskie 4, 00-478 Warsaw, Poland
                 \ \ email: gwiktoro@astrouw.edu.pl}
\altaffiltext{2}{Center for Gravitational Wave Astronomy, University of 
                 Texas at Brownsville, Brownsville, TX 78520, USA}
\altaffiltext{3}{Department of Physics, Texas Tech University, 
                 Box 41051, Lubbock TX 79409-1051}

\begin{abstract}{
There are $19$ confirmed BH binaries in the Galaxy. $16$ of them are X-ray 
transients hosting a $\sim5-15\msun$ black hole (BH) and a Roche lobe overflowing 
low-mass companion. Companion masses are found mostly in $0.1-1\msun$ mass 
range with peak at $0.6\msun$. The formation of these systems is believed to 
involve a common envelope phase, initiated by a BH progenitor, expected 
to be a massive star $>20\msun$. It was realized that it may be very
problematic for a low-mass companion to eject a massive envelope of the black 
hole progenitor. It invoked suggestions that an intermediate-mass companion 
ejects the envelope, and then is shredded by the Roche lobe overflow to its
current low-mass. But this creates another issue; a temperature mismatch 
between hot models and the observed cool low-mass donors. Finally, the main 
driver of Roche lobe overflow that is believed to be magnetic braking does not 
seem to follow any theoretically calculated models. Number of ideas were put 
forward to explain various parts of this conundrum; pre-main sequence donor 
nature, alternative approach to magnetic braking and common envelope energy was 
revisited. We test various proposals and models to show that no overall solution 
exists so far. We argue that common envelope physics is not crucial in the 
understanding of Galactic BH transients. Our failure most likely indicates that 
either the current evolutionary models for low-mass stars and magnetic braking 
are not realistic or that the intrinsic population of BH transients is 
quite different from the observed one.
}
\end{abstract}

\keywords{
binaries: close --- stars: evolution --- X-rays: binaries   
}

\section{Introduction}

Nearly all confirmed Galactic BHs reside in X-ray Transient systems (BHXRTs). 
These are close interacting binaries comprising a BH and a low mass 
companion. Companions are mostly main sequence stars with a few exceptions of evolved 
stars (see Table \ref{tab:bhxrbs}). In this work we adopt the terminology where the star 
which is heavier on Zero Age Main Sequence (ZAMS) and therefore is the one to 
form the compact object is called a primary, whereas the other star is called 
a secondary.

Roche lobe overflow mass transfer leads to the formation of accretion disk around black 
hole (e.g., \citet{Shakura1973}), which is the source of X-ray radiation. BHXRTs 
spend most of the time in the so-called quiescent state when the X-ray emission is 
negligible in comparison to the optical emission of the secondary. The distinctive 
feature are X-ray outbursts with the brightness increase of a few ($\sim 3-5$) 
orders of magnitude. During the outbursts X-ray luminosity significantly exceeds
optical  luminosity ($L_x/L_{opt}\gg1$). In few systems the recurrences have been 
observed (e.g., \citet{McClintock2006}). Disk instability connected to ionization 
of hydrogen atoms is claimed to be responsible for the outbursts (e.g., \citet{King1996},
\citet{Dubus1999}, \citet{Lasota2001}).
It is expected that transient behavior appears only for very low mass transfer rates. 
This naturally agrees with the low mass companions in the observed population.     
Alternatively, it was proposed that two-partial accretion disk is responsible for 
transient behaviour \citet{Menou1999}. The inner Advection Dominated Accretion Flow (ADAF) part
is responsible for the emission in quiescence, whereas the outer thin disk part may
be, under some conditions, unstable and, therefore, responsible for the outbursts.

Black holes are predicted to form from massive stars ($>20-40 \msun$, e.g., 
\citet{Fryer2012}). Since massive stars have large radii ($>10-20 \rsun$, e.g., 
\citet{Hurley2000}), initial binary orbits need to have rather large separations. 
Intensive wind mass loss (e.g., \citet{Vink2013}) from a massive star may lead to a 
significant increase of the binary orbit. One may expect a typical orbital separation 
to be rather large ($\gtrsim 100 \rsun$) during the early evolution leading to the 
formation of BHXRT. Since the observed BHXRTs are Roche lobe overflow binaries with 
low mass companions (typically small radii), the orbital separations of the
observed systems are rather small ($\lesssim 10 \rsun$). There must exist a
process that leads to the significant orbital decay of the BHXRT progenitor. 
All Galactic BHXRTs are found in the isolated environment (in the Galactic field) 
with no apparent nearby dense clusters or any extra stellar components. 
In isolation, common envelope (CE) phase \citep{Paczynski1976} is the best known 
mechanism for decreasing the orbit and the formation of close and interacting 
binaries. 

The evolution leading to the formation of BHXRT may start with the detached binary 
of rather extreme mass ratio. The massive star evolves off main sequence, increases 
its radius and initiates the Roche lobe overflow. The mass transfer proceeds
on dynamical timescale and leads to the formation of CE (i.e., primary envelope 
engulfs entire binary system). After CE, a massive and compact core of the
primary emerges on a close orbit with the unaffected companion. 
The primary core rapidly evolves towards core collapse and the BH
formation. At this stage the orbital separation needs to be so small that 
evolutionary expansion of a companion and/or combined action of magnetic braking, 
tides and emission of gravitational radiation from the system allows for the Roche
lobe overflow. Companion may start at arbitrary mass and then become low
mass star by mass loss during the Roche lobe overflow phase. The mass transfer 
from low mass companion to a BH proceeds at the very low rate leading to 
the disk instability and transient behavior.  

There is a number of issues with this picture. As noted early on by number of authors 
(\citet{Podsiadlowski1995}; \citet{Portegies1997}; 
\citet{Kalogera1999}) and then later supported by more recent studies (e.g., \citet{Podsiadlowski2003}; \citet{Kiel2006}; \citet{Yungelson2008}) it was recognized that 
survival of common envelope phase is rather challenging energetically. Ejection of 
massive envelope of a BH progenitor by a low-mass star proves rather difficult if not 
impossible. This issue is approached with extra sources of energy (other than 
orbital) that may help unbind the massive envelope. It usually translates to invoke 
very high CE ejection efficiencies. 

For example \citet{Yungelson2008} have found a need for 
$\alpha=10$-$40$\footnotetext{Actually, these authors list only $\alpha \times \lambda 
= 0.5$-$2$. We have used recent estimates of binding energy for massive BH progenitors 
($\lambda=0.05$; \citet{Xu2010b}) to estimate their $\alpha$. Definitions of 
$\alpha$ and $\lambda$ are given in Section \ref{sec:mstd}}, and have justified such high values 
with similar findings in other evolutionary studies. It needs to be noted that their 
findings were obtained with rather high wind mass loss rates for O stars as compared 
with the most updated theoretical predictions (\citet{Vink2001}; although see \citet{Eldridge2008}). High wind mass loss helps to remove part of the envelope before CE is 
encountered. Therefore, the already very high $\alpha$ values found by \citet{Yungelson2008} would increase if reduced wind mass loss rates were applied. 

\citet{Podsiadlowski2010} speculated that possibly hydrogen-rich material from the 
inspiraling low-mass main sequence star can be injected into the helium-burning shell 
of the evolved BH progenitor. In fact, if nuclear energy can help eject the
envelope it appears that $1-3\msun$ donors can easily survive CE and form close systems 
with BHs.  However, the subsequent evolution through RLOF and transient phase 
(as we will show in our study) will lead to the depletion of systems with mass 
($\sim0.6\msun$) where most observed Galactic transients are found and will favor 
systems with higher mass ($\sim1\msun$). However, the proposed scenario for the 
efficient CE ejection may eliminate any potential issue with the low number of
predicted systems in future modeling.
 
\citet{Ivanova2011} provided a refined binding energy estimates. Based on 
considerations of envelope internal behavior and energy they argued that the binding 
energy is about factor of $\sim 2-5$ smaller than usually estimated and suggested 
that this may help to form close BH systems with low mass companions. As we will show 
in our calculations this helps to increase the number of Galactic BH transients, 
however the donor mass distribution peaks at $\sim 1\msun$ rather than at observed 
$\sim 0.6\msun$. 

\citet{Justham2006} argued for another line of approach to this issue with a 
proposal that it is intermediate-mass companions that eject the common envelope. It 
was proposed that later in the evolution these companions are mass depleted by the 
RLOF onto a BH and they become low mass stars. However, this created another 
problematic issue. Such stripped stars would be hotter than low-mass stars observed 
in the Galactic BH transients. 

As \citet{Justham2006} proposed magnetic braking for intermediate-mass donors 
(namely magnetic Ap and Bp stars) to be a major angular momentum loss
mechanism driving system evolution, \citet{Chen2006} proposed a different way of 
angular momentum loss from intermediate-mass stars. During the RLOF some of the 
matter lost from the donor may not reach the primary but form a circumbinary
 (CB) disk. This can potentially be the source of tidal torques, which
may result in the effective angular momentum drain. However, this proposal also 
suffers from the temperature mismatch problem.

\citet{Ivanova2006} delivered a potential solution of this issue. It was proposed that it 
is pre-main sequence stars donors feeding BHs in the Galactic transients. Pre-main 
sequence donors could have an intermediate-mass, but because they are cooler than 
main sequence stars, the temperature issue would be avoided. However, none of the 
Galactic BH transients are associated with or found nearby star-forming regions. 

Finally, \citet{Yungelson2006} and \citet{Yungelson2008} pointed out that only 
the models with reduced or no magnetic braking for the evolution through BH transient 
phase (during ongoing RLOF) seem to be encouragingly close to observations. 
Unfortunately, recent observations of orbital decay in BH transients seem to indicate 
very high magnetic braking rate \citep{Gonzalez2014}. Despite these 
observations we have tested this idea, to show that it produces an extra evolutionary 
problem. 

The population of known X-ray binaries have a significant impact on our understanding 
of evolution of both high-mass and low-mass stars. For example, the NS and BH mass 
spectrum, with the characteristic lack of compact objects within mass range $2$--$5\msun$ 
can be qualitatively explained with the specific model of supernova explosion 
\citep{Belczynski2011} or it may be an observational bias in BH mass measurements 
\citep{Kreidberg2012}. The companion mass spectrum of Galactic BHXRTs still awaits to be 
explained. The companion mass distribution peaks at $0.6\msun$ and majority companions 
are found with mass below $1\msun$ (see Table \ref{tab:bhxrbs}. As pointed out above 
it seems rather difficult to explain the existence of such systems with the standard 
evolutionary scenarios. 
We reexamine this issue to show that {\em (i)} common envelope is not a crucial issue 
in the formation of BH transients as we can produce them without invoking unrealistically 
high common envelope efficiencies, {\em (ii)} magnetic braking efficiency 
plays a crucial role in shaping the donor mass distribution and that {\em (iii)} we 
cannot reproduce the observed population in any of our simulations. 

In the next section (\s\ref{sec:modeling}) we 
briefly describe our population synthesis code {\tt StarTrack} and several CE 
models are presented. In the following section (\s\ref{sec:results}) 
we discuss our predicted companion mass distributions and compare them with 
observations. Next section (\s\ref{sec:selection}) contains discussion of selection effects on 
the X-ray binary population. Finally, our conclusions are presented in the last 
part (\s\ref{sec:conclusions}).

\section{Modeling}\label{sec:modeling}

\subsection{Population synthesis model}\label{sec:startrack}

We have made use of the {\tt StarTrack} population synthesis code. 
It was created with particular attention put on the evolution of     
massive stars, as they can produce neutron stars (NS) and  BHs. 
The single stellar models were adopted from \citet{Hurley2000}, who 
focused mostly on evolution of low mass stars and white dwarf 
formation. A thorough description of the code may be found at 
\citet{Belczynski2002b} and \citet{Belczynski2008}.

Evolution leading to the formation of BH transients involves two distinctive 
phases: CE inspiral and BH formation. The approach to CE is described in the
following subsections, here we give brief description of BH formation. We
use a core collapse/supernova model that is based on neutrino supported
convective explosion engine with rapid explosion development \citep{Fryer2012}.
This model results, with qualitatively reasonable compact object mass 
spectrum. Most NSs are found with mass $\sim 1.3$--$1.4\msun$ but with a long tail 
extending to over $2\msun$. Galactic BHs (solar metallicity models) are found
with mass in range $5$--$15\msun$. Note that this reproduces the observed
mass gap between NSs and BHs (\citet{Belczynski2012}; although see \citet{Kreidberg2012}). For NSs we adopt high natal kicks during supernova explosion
derived from observations of pulsar velocities in Galaxy; the Maxwellian 
distribution of kicks with $\sigma=265$km s$^{-1}$ \citep{Hobbs2005}. For BHs we
use the same distribution but with lowered $\sigma$, as significant fraction
of the mass ejected during SN explosion may falls back onto the BH.
As a result, natal kicks for our BHs are small or negligible. This
is representative for kicks originating from asymmetries in mass ejection
during SNa explosion. It seems that our model is consistent with majority of
BH natal kick estimates \citep{Belczynski2012}.   

For massive primaries we have adopted a steep initial mass function with power-law 
exponent of $-2.7$ \citep{Kroupa1993}. We have chosen the primary mass in range $6$--$150\msun$ because 
less-massive stars are not able to evolve into a BH. Typically, in
our single stellar models only stars with ZAMS mass greater than $20\msun$
may form BHs. However, much lower mass limit was adopted as binary 
interactions (e.g., mass transfer) may effectively lower this BH threshold
formation mass. The secondary mass has been taken from a wider distribution
$0.08$--$150\msun$ in such a way that the ratio of secondary to primary 
mass on ZAMS has a uniform distribution and is always smaller than $1$ \citep{Kobulnicky2006}. The 
distributions of initial separations ($a$) and eccentricities ($e$) are 
proportional to $1/a$ \citep{Abt1983} and $e$ \citep{Duquennoy1991} respectively. Separations were limited to be 
greater than two times the sum of stars radii as to allow the formation of
a binary in the first place and smaller than $10^5\rsun$. Whereas eccentricity 
is in range from $0$ to $1$. We have assumed binarity of $50\%$ ($2/3$ stars in
binaries). Such parameters are consistent with the observed Galactic binary 
population.  

Recently, \citet{Sana2012} and \citet{Sana2013} have obtained moderately different 
initial distribution for massive O-type stars. In particular, it was found
that there are somewhat more close binaries than implied by flat (in
logarithm) separation distribution, also the binarity was found at higher
level $\sim 70\%$ and it was concluded that massive O binaries are less eccentric 
than what we have adopted (i.e., they follow distribution $\propto e^{-0.42 \pm 0.17}$). 
It is not at all clear that these results are applicable for the progenitors
of BH transients; binaries with O star and K-M dwarf as the lowest mass
object in the \citet{Sana2012} observational sample was a $16\msun$ star.   
If these distributions were applied across entire binary mass range we would
expect a moderate increase in a number of BH transients. 

We have evolved $2\times10^7$ binaries starting from ZAMS. We adopt a $10\gyr$ 
time limit for the evolution of our synthetic binaries, which corresponds to 
the age of Galactic disk. We have also assumed that the star formation in
the disk was constant throughout the last $10\gyr$. For each binary we note
the time and duration of an X-ray transient phase with BH accretor (if 
encountered). Then we estimate the probability of finding a given binary in 
this phase at the current time. This way we can study physical properties 
of synthetic BHXRTs and compare them with the observed population.  

In the following subsections we describe in detail how we treat and test 
various physical processes that may play an important role in the formation 
and subsequent evolution of BHXRTs.

\subsection{Standard energy CE model}\label{sec:mstd}

The orbital energy in a binary system can be expressed with the formula
\(E_{orb}=-\frac{Gm_1m_2}{2a}\)
where $G$ is the gravitational constant, $m_1$ and $m_2$ are star masses,
and $a$ is the average distance between the stars (separation). The loss of
binary energy can be calculated as the difference between its
energy before and after the CE phase. 
\[\Delta
E=E_{pre}-E_{post}=-\frac{Gm_{1,pre}m_{2}}{2a_{pre}}+\frac{Gm_{1,post}m_{2}}{2a_{post}},\]

where $m_{1,pre}$ and $m_{1,post}$ are the masses of primary star before and
after losing its envelope, respectively. Similarly, $a_{pre}$ and $a_{post}$ are separations
before and after the CE. We assume here that the mass of the low mass main
sequence companion remains unchanged (no significant accretion). 

The energy needed to unbind the envelope from the system is the combination of
its internal energy and potential energy, but for the use of population
synthesis code it is usually simplified to  
\begin{equation}\label{eqn:lambda}
    E_{bind}=\frac{Gm_{1,pre}m_{1,env}}{\lambda R_{1,rl}}
\end{equation}
where $m_{1,env}=m_{1,pre}-m_{1,post}$ is the mass of the envelope, $R_{1,rl}$ 
is the Roche lobe radius of the primary star, and $\lambda$ is a parameter 
describing the binding energy of the envelope \citep{deKool1990}. 
We use physical estimates of $\lambda$ for stars of different size, mass and
evolutionary state as calculated by \citet{Xu2010}. In this calculation the 
binding energy was estimated based either on gravitational energy only, or
it was decreased by its internal energy. Here we use the average of the two. 
The typical values for BH progenitors are of the order $\lambda=0.05$, and 
they do not ever rise significantly above $\lambda=0.2$ \citep{Dominik2012}.

It is noted that the binding energy scaling was systematically studied by other 
groups (\citet{Dewi2001}, Northwestern). In particular, \citet{Dewi2001} 
noted that for stars with $M\leq20\msun$ (e.g., NS progenitors) the $\lambda$ may 
vary by large factors (almost $2$ orders of magnitude) depending on the adopted 
core definition. They quoted values in range $\lambda=0.05$--$3.5$ for their
highest mass model, a $20\msun$ star. Note that for our BH progenitors we
use the lowest of their value, increasing binding energy and making CE ejection 
vary hard. Especially for systems with low mass companions. In other words 
we are providing a conservative approach to the CE survival and BH transient
formation. Recently, it was demonstrated that for more massive stars (e.g., 
BH progenitors) the core definition does not change significantly the
envelope binding energy \citep{Wong2013}.

The energy transfer from the orbit to the envelope may not be ideal so the 
parameter $\alpha$ was introduced. It describes what fraction of the binary
orbital energy is effectively used for unbinding the envelope. We adopt
$\alpha=1$ as to allow for the easiest possible envelope ejection by a low
mass star. The final equation on energy transfer can be written as
\begin{equation}\label{eqn:a}
    \alpha\left\{\frac{Gm_{1,post}m_{2}}{2a_{post}}-\frac{Gm_{1,pre}m_{2}}{2a_{pre}}\right\}=\frac{Gm_{1,pre}m_{1,env}}{\lambda R_{1,rl}}
\end{equation}

This readily provides an estimate of the $a_{post}$ from the parameters before the CE phase 
as the $m_{1,post}$ is assumed to be equal the core mass of the donor, which is 
given by evolutionary models \citep{Hurley2000} employed in the {\tt StarTrack} 
code.

The above energy balance provides only simplified picture of CE as
introduced early on by \citet{Webbink1984} and its shortcomings were recently
discussed by \citet{Ivanova2013}. 
We will use it as our reference model to {\em (i)} test whether any
improvement was achieved with recent population synthesis updates since
original work of \citet{Podsiadlowski2003}, and to {\em (ii)} contrast its
predictions with other available models that we test in this study.

\subsection{Enthalpy CE model}\label{sec:mivn}

\cite{Ivanova2011} proposed a modified approach to the standard energy CE model. 
They come up with a conclusion that the sum of gravitational and thermal energies 
may not be the correct value for binding energy. Showing that even positive total 
energy could describe a stable system, they suggested to use enthalpy instead of 
energy as a fundamental parameter. They base their idea on the fact that transfer 
of the energy from companion to envelope will lead not only to the rise of the 
thermal kinetic energy of the gas but also to the expansion of the envelope. This 
expansion will result in work done by envelope. As a result we may expect that a 
smaller energy will be needed to unbind the envelope. This concept can be 
incorporated into our calculations by appropriate multiplication factor 
($f_{\lambda}$) to the $\lambda$ value in Equation \ref{eqn:lambda}. \cite{Ivanova2011} 
estimated it in the range $f_\lambda=2\div5$. 

We have checked the impact of this approach on BHXRT companion mass distribution 
for the most extreme case, $f_\lambda=5$, that allows for the easiest ejection of 
the envelope. In principle, it is expected that typically companions of a lower mass
will be found in this model as compared with standard energy model. 

\subsection{Angular momentum CE model}\label{sec:mnel}

This CE model incorporates the angular momentum balance. The idea was first
introduced by \citet{Paczynski1967}, and recently employed by \cite{Nelemans2000} 
in population synthesis studies.  
A linear loss of angular momentum with matter ejected from binary during the 
CE phase leads to
\[
\gamma=\frac{\Delta J}{J}/\frac{\Delta M}{M}
\]
where $\gamma$ represents a constant ratio of angular momentum $J$ loss to mass 
loss $M$, both referring to the entire binary. This can be translated into the 
final separation after CE in the form
\[
a_f=a_i \left(\frac{m_{1,pre}m_{2,pre}}{m_{1,post}m_{2,post}}\right)^2 \frac{M_{post}}{M_{pre}}\left( 1-\gamma \frac{M_{pre}-M_{post}}{M_{pre}} \right)^2,
\]
where $M_{post}=m_{1,post}+m_{2,post}$ and $M_{pre}=m_{1,pre}+m_{2,pre}$ are the 
total masses of the system after and before the CE phase. In this model, as in 
previous ones, we assume that the entire envelope is being expelled from the 
binary, leading either to a formation of a close system or merger of both
interacting binary components. 

For comparison we list below calculated final separation for the three employed
CE models obtained for the exactly same system entering the CE phase. At
the beginning of CE phase we have a core Helium burning star with mass $23.4
\msun$ (with He core of $15 \msun$) and a $3.6 \msun$ main sequence companion
on the circular orbit with separation $a=4000\rsun$. After CE we obtain
separation $15\rsun$ (energy balance), $73\rsun$ (enthalpy) and $1921\rsun$
(angular momentum).

\subsection{Enhanced mass transfer model}\label{sec:mill}

Additionally, we have checked the idea of enhancing the mass loss from the
secondary due to illumination of the Roche lobe filling stellar envelope by the
high energy radiation. This radiation is produced in the accretion disk around 
the BH and is radiated in all directions. Some part of it will, therefor, 
be captured by the secondary's envelope. The illuminated envelope may
increase its temperature and size and this in turn may potentially lead to
the increased mass transfer rate. 

The exact calculation of envelope response to illumination is beyond the scope 
of this paper. Instead, we adopt an increase factor $f_{\rm ill}$ to the mass
transfer rate from low mass companion to a BH. The mass transfer is
first obtained for single stellar models of \citet{Hurley2000} by calculation 
of a star and its Roche lobe response to the mass loss with inclusion of
magnetic braking \citep{Belczynski2008}. Then to mimic a potential mass loss 
increase we multiply it by several values of $f_{\rm ill}=2, 5, 10$.  

Since the past, and as we will see the current models, are generating typically
companions of too high mass in BHXRTs, the expectation is that illumination will 
help to bring models to match the observations.

\subsection{Low maximum NS mass model}

In our calculations we have adopted the maximum NS mass
at the value $M_{max,NS}=3\msun$. Here, we let this limiting value to
decrease down to $M_{max,NS}=2\msun$. This may have an important effect on
BHXRT populations as there is a possibility that a BH in a close
binary system may form via accretion induced collapse (AIC) of a NS. 
The lower the maximum NS mass the easier it is to form BHs via AIC.  
Observationally, the adopted here low value is consistent with the two most 
massive known NSs ($1.97\pm0.04$ \citep{Demorest2010}, $2.01\pm0.04$ \citep{Antoniadis2013}). 
Theoretical calculations give maximum NS mass in the range $2.0$--$3.2\msun$
(e.g., \citet{Haensel2007}, \citet{Chamel2013} and 
\citet{Kiziltan2013}). 

As we can see from Table \ref{tab:bhxrbs}, the lowest mass BHs
are above $\sim 5\msun$, although the mass estimates may be
observationally overestimated (see \citet{Cantrell2010} for a discussion of how past work has been flawed and how such
problems may be fixed in the future; see also
\citep{Kreidberg2012} for a discussion of how this effect impacts BH populations on the whole).

\section{Results}\label{sec:results}

In this section we test various models, assumptions and proposals from the
literature that were put forward as potentially important in the formation
and evolution of BH transients. 

We consider only binaries with companions less massive than $2\msun$.
In Table \ref{tab:bhxrbs} we list Galactic BH binaries. The most known BH transients have
companion masses below $2\msun$. This agrees well with the theoretical predictions that 
require low mass transfer rates and thus low mass donors for transient behavior to appear
(e.g., \citet{Dubus1999}, \citet{Menou2002}).

There are two exceptions found in our table: XTE J1819-254 and
4U1543-47. The XTE system has reported companion mass
$5.5-8.1\msun$. However, it appears that the companion is enriched
with CNO burning products and this leads to the increased mass
transfer rates allowed for the transient behavior
(e.g., \citet{Dubus1999}). Furthermore, the system's outbursts do not
appear like the outbursts of the other X-ray binaries -- instead, they
have been dramatically shorter, and dramatically brighter than those
of other X-ray transients (see e.g. \citet{Hjellming2000}).  
In our analysis we consider only binaries with H-rich envelopes
and typical solar-like composition. This system is beyond the scope of
our study.  The 4U~1543-47 binary is close to our imposed limit with
companion mass estimated to be in range $2.3\div2.6$. However, the
mass estimate is a subject to several uncertainties. \citet{Orosz1998}
discussed the possibility that the A2V star may not be the part of XRB
and be either the background star or the outer part of the triple
system.  Despite the fact that comparison of mass transfer rate for
such massive donors allows for the transient behavior we limit our
study to stars below $2\msun$. Note that in each model we find a small
number of transients with masses above $2\msun$. Inclusion of this one
particular system in our analysis would not change any of our
conclusions.

\subsection{CE survival with low mass companions}

We note that even stars as low as $0.5-1\msun$ may successfully eject an envelope 
of a very massive stars (i.e., BH progenitor; $M_{\rm zams}=20$--$150\msun$).
And this result is obtained for very restrictive values of $\lambda$ that we
have applied in our calculations. For massive stars with $M_{\rm zams}<75\msun$
we employ $\lambda=0.05$, while for higher mass stars $\lambda=0.4$ \citep{Xu2010b}.
It means that for a typical BH progenitor with $M_{\rm zams}=30-40\msun$ we multiply 
binding energy by factor of $20$ ($1/\lambda$) making the envelope very hard to be 
ejected. 

There are two ways (or their combination) to overcome the ejection problem. First, 
if CE is initiated on a very large orbit then there is a lot of orbital contraction 
available (i.e., orbital energy reservoir is large) for successful ejection of massive 
envelope even by low mass stars. Second, if CE is initiated late in terms of evolution 
of massive star it means that massive star lost most of its envelope in stellar winds. 
That means that not much orbital energy is required and successful envelope ejection
may be achieved with low mass star. Below we give two examples of such ejections. 

The binary starts with $100\msun + 1\msun$ components. At the time of CE onset the 
primary has just begun core He-burning ($\sim1-2\%$ into this phase that lasts about 
$0.25$ Myr for this star). At this time its mass was decreased by winds down to $60\msun$, 
with $25\msun$ He-rich  core and massive $35\msun$ H-rich envelope. Pre-CE orbital
separation is $a=4000$--$5000\rsun$ and star radius is $\sim 3000\rsun$. After energy 
balance application with CE efficiency $\alpha=1$ and binding energy scaling of 
$\lambda=0.4$ (see Equation \ref{eqn:a}) the post-CE separation is $5-10\rsun$; wide enough to accommodate 
$25\msun$ WR star and its $1\msun$ companion, and close enough to allow for
onset of RLOF from the low mass star within the Hubble time. 

The binary starts with $30\msun + 0.5\msun$ components. At the time of CE onset the 
primary is just about to finish core He-burning ($\sim80\%$ into this phase that lasts about 
$0.5$ Myr for this star). At this time its mass was decreased by winds down to $13\msun$, 
with $10\msun$ He-rich core and low mass $3\msun$ H-rich envelope. Pre-CE orbital
separation is $a=2000$--$3000\rsun$ and star radius is $\sim 1500\rsun$. After energy 
balance application with CE efficiency $\alpha=1$ and binding energy scaling of 
$\lambda=0.05$ (see Equation~\ref{eqn:a}) the post-CE separation is $\sim5\rsun$; wide enough to 
accommodate $10\msun$ WR star and its $0.5\msun$ companion, and close enough to allow for
onset of RLOF from the low mass star within the Hubble time.

\subsection{Standard model}\label{sec:rstd}

\begin{figure}
\begin{center}
    \includegraphics[scale=0.6]{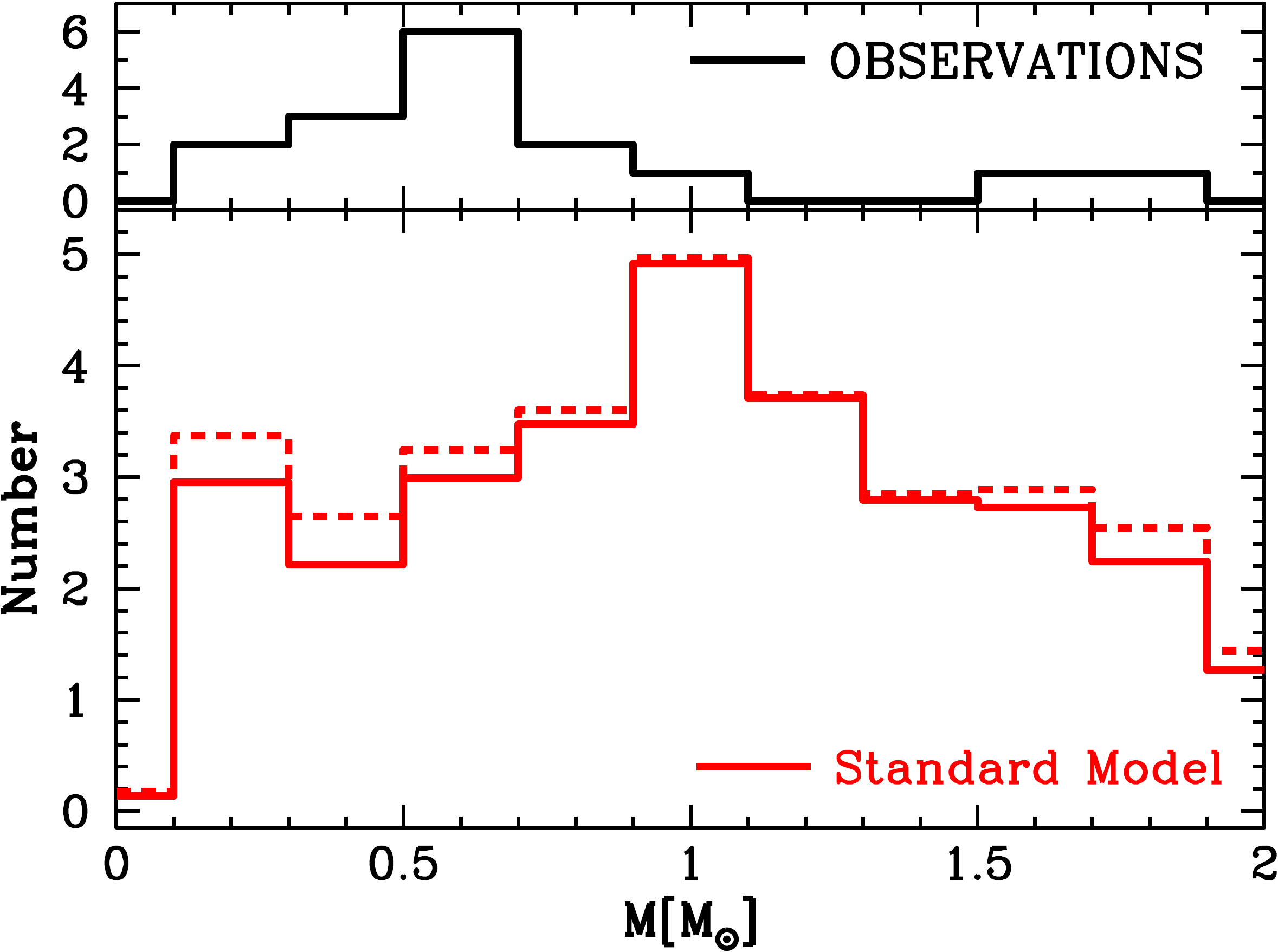}
\end{center}
    \caption{
Top panel: observationally estimated donor masses for 16 Galactic BH transients with
low mass companions. 
Bottom panel: distribution of donor mass for BH transient population obtained with 
our standard evolutionary scenario that employs energy balance for common envelope 
(see Section \ref{sec:mstd}). Solid line shows predicted current Galactic population 
of BH transients, while dashed line shows both transient and persistent systems.   
We find about $30$ systems in the mass range $0-2\msun$. Observations peak at 
$\sim0.6\msun$, whereas the simulation peak at $\sim1\msun$. 
    }
    \label{fig:std}
\end{figure}

Figure \ref{fig:std} shows the predicted mass distribution for companion
stars in Galactic BHXRTs. The distribution covers a broad range of masses,
with significant numbers of stars in a mass range $0.1-2\msun$. There is a
clear peak in the distribution at $\sim 1\msun$, and a smaller one at
$\sim 0.2\msun$. The predicted distribution is quite different from the
observed one that has a peak at $\sim 0.6\msun$. 

Binaries responsible for the main peak of the predicted distribution
consist of a rather massive primary $\sim 30-40\msun$ and a low mass
secondary $\sim 1-1.2\msun$.  Rapid evolution of primary leads to the
Roche lobe overflow while it is core He burning. After significant
orbital decay ($a=3000\rsun$; energy balance; $a=6\rsun$), the close
binary emerges out of CE. Massive He core collapses to form a $\sim
7-10 \msun$ BH at about $6\myr$ since ZAMS. For such a massive star we
assume the direct BH formation with no (or weak) supernova explosion
and the orbit remains very close $a\approx8\rsun$ (no or small natal
kick). Nuclear evolution on main sequence increases the radius of
secondary.  At the same time magnetic braking (MB) and gravitational
radiation (GR) keep decreasing orbital separation over next $4\gyr$
until secondary fills its Roche lobe at $a\approx5\rsun$.  The
secondary initiates stable Roche lobe overflow (X-ray phase begins)
during its main sequence evolution at the low mass transfer rate
($\dot{M} \approx 8\times10^{-11} \msy$).  At such a low mass
transfer rate the system is a subject to disk instability and becomes
a transient for several \gyr s. For this particular binary configuration
the critical mass transfer-rate below which the system shows transient
behavior is $\dot{M}_{\rm disk} \approx
10^{-10} \msy$ \citep{Dubus1999}.  Due to such slow mass transfer
rate, donor mass changes very slowly. It means that there is a high
probability for the system to be discovered with secondary close to
its initial mass. This explains the main peak in BH XRB companion mass
distribution at $\sim 1-1.2\msun$.
 
During stable mass transfer onto BH we assume that mass transfer is
limited to Eddington critical rate ($\dot{M}_{\rm Edd} \approx
10^{-7} \msy$) therefore this system follows a fully conservative
mass transfer mode. In such a case mass transfer from lower mass star
to more massive BH leads to the orbital expansion. However, this is
counter-balanced by combined angular momentum loss via GR and MB and
the orbit decays. With the decreasing semi-major axis the strength of
GR increases and the rotation of the donor speeds up increasing the
strength of MB \citep{Ivanova2003}. This leads to the increase in mass
transfer rate ($\dot{M} \approx 10^{-10}-10^{-9} \msy$) when mass
of the companion drops below $0.9-0.7\msun$ and separation decreases
to $a=2-3\rsun$.  This has two consequences. First, the donor mass
depletion accelerates so the system contributes less and less to the
companion mass distribution in the decreasing mass bins. We note the
gradual decrease in number of systems leftward of $1\msun$. Second,
for some systems the binary parameters are such that mass transfer may
accelerate over $\dot{M}_{\rm disk}$ and system becomes a persistent
XRB causing further depletion in our distribution as only transient
mass distribution is of interest here (the observations shown in Figure \ref{fig:std}
include only transient BH XRBs).  The intensive mass transfer rather
quickly ($0.5$\gyr) depletes donor mass below $0.35\msun$ and at this
point the star becomes fully convective and MB turns off. Mass
transfer slows down ($\dot{M} \approx 5\times10^{-11} \msy$) and
donor mass slowly drops down to hydrogen burning limit ($\sim
0.1 \msun$) when we stop calculations. This last phase lasts $\sim
1-2\gyr$ and due to slow evolution in mass transfer these systems form
a secondary peak in donor mass distribution at $0.2\msun$.
Note that mass transfer rates calculated with our population synthesis 
code {\tt StarTrack} are in agreement with ones inferred from 
observations. For example, the mass transfer rates for BH transients 
were estimated at the level of $\gtrsim 10^{-10}\msy$ by \citet{King1988} 
and $\sim 10^{-10}\msy$ by \citet{Justham2006}.

Systems with secondaries of lower mass than in the above example
follow very similar evolution. They enter the mass distribution at
lower mass and move across it in a similar fashion as we assume
effective de-rejuvenation of mass losing stars.  The parameters (i.e.,
radius and luminosity setting mass transfer rate) of a star that is
$0.9\msun$ and on main sequence are not drastically different whether
this star was born with $0.9\msun$ or was born as $1.1\msun$ star and
then reduced to $0.9\msun$ by RLOF \citep{Hurley2000}.

Systems with higher mass secondaries evolve quite differently than in
the above example. Primary is more massive than $\sim40\msun$, while
secondary is typically $\sim 3-4\msun$ star. Rapid evolution of
primary leads to the CE while it is core He burning. After significant
orbital decay, the close binary emerges out of CE.  Massive He core
collapses to form a $\sim10\msun$ BH at about $5\myr$ since ZAMS.  For
such a massive star we assume the direct BH formation with no
supernova explosion.  The companion initiates RLOF during its main
sequence evolution due to its nuclear expansion and GR emission at
$\sim 100 \myr$.  Stable mass transfer from a rather massive main sequence (MS) star
begins removing mass from the secondary at moderate rate
($\dot{M} \approx 5\times10^{-9} \msy$). System appears as
transient XRB at this phase. During this phase orbital separation
increases from $a=4\rsun$ to $a=5\rsun$ (conservative mass transfer
from less massive to more massive binary component, with no
intervening MB). This phase lasts for $\sim 1-2\gyr$ until the donor
becomes low mass MS star ($M_b<1.25\msun$) and develops a convective
envelope. Since mass transfer rate is rather constant during this
phase the resulting mass distribution is flat above $1.25\msun$. Below
this mass the donor is a subject to MB and mass transfer significantly
increases ($\dot{M} \approx 2.5\times10^{-7} \msy$). This phase
lasts only about $1$Myr but donor drops in mass to $M_b=1\msun$ and
becomes a HG star. Orbit has expanded (high mass transfer rate wins
over MB and GR) to $a=7\rsun$. At this short phase the system becomes
a persistent XRB and disappears from our population of interest. These
systems do not contribute to the main peak of the distribution.  A HG
star, that has formed from de-rejuvenated $3-4\msun$ MS star, has
temperature above $\sim 5000K$ and has most likely radiative envelope
(e.g., \citet{Belczynski2008}).  For such a star MB turns off and mass
transfer slows significantly down ($\dot{M}
\approx 10^{-10} \msy$). The system re-appears as transient on the left side of 
the main peak. At lower mass ($M_b\lesssim0.9-0.8\msun$) donor becomes red giant with 
convective envelope, MB turns on again increasing mass transfer rate ($\dot{M} 
\approx 10^{-9}\msy$). The combined effects of MB and GR overcome such moderate 
mass transfer rate and lead to a decrease of orbital separation. On
the other hand, mass loss from convective red giant envelope along
with its evolutionary expansion leads to donor's radius increase and
when donor's radius exceeds its Roche lobe radius by factor of $2$ we
stop calculations.

Our condition to terminate calculations is somewhat arbitrary. In one
alternative simulation we have relaxed this condition. MB (convective
red giant) and GR (very close system $a\sim3-4\rsun$) decrease the
semi-major axis. At the same time, a low mass red giant donor ($M_{\rm
b} \sim 0.6 \msun$) responds with expansion to mass loss. Therefore,
we have a runaway situation that in a short time leads to a CE and
termination of BHXRT phase.

\subsection{Enthalpy model}\label{sec:rivn}

Figure \ref{fig:ivn} shows the predicted companion mass distribution
for increased $\lambda$ value ($f_\lambda=5$). The distribution is
very similar in shape to standard model distribution with main peak at
$\sim1 \msun$ and secondary peak at $\sim 0.2 \msun$. Despite the
expectation to form BHXRTs with lower mass companions (due to easier
envelope ejection) we do not find significant improvement. The
predicted companion mass distribution is still in tension with the
available observations.

We also note the increase in number of BHXRTs in this model as
compared to the standard model. This is the combined result of initial
distribution of binary separations and CE filtering. We start with the
(statistically) same population of binaries in each model. For a given
system to become a BHXRT a very specific post-CE separation is
required ($a \sim 5-10\rsun$). It means that for similar binary
component masses in the standard model pre-CE (and thus initial ZAMS)
separation is larger than in the enthalpy model (in which the envelope
binding energy is smaller). In each model there is more binaries with
small separations as compared to number of systems with large
separations. Initial separations are drawn from distribution flat in
logarithm ($\propto 1/a$).

The obvious counter-argument to the above reasoning follows from the
IMF. There are more low mass stars (i.e. $0.5-0.7\msun$) forming the
peak of observed distribution than high mass stars produced in the
peak of our model distributions ($1-1.2\msun$).  There are three major
counter-acting factors to the IMF argument.
 
{\em (i)} Stars below $1\msun$ almost do not expand within Hubble
time. It means that the range of post-CE separations is rather narrow
for the system with such low mass companion to later enter RLOF and
become BHXRT. The onset of RLOF is further hindered by the small mass
of one component (donor) that leads to less efficient orbital angular
momentum loss via GR and MB.

{\em (ii)} As in standard model (main example of evolution) systems
with main sequence donors below $1\msun$ evolve relatively quickly
toward very low mass therefore their contribution to the companion
mass distribution of BHXRTs is small.  The rapid evolution follows
from high mass transfer rates caused by combined effects of GR and MB
enhanced by small separations required for this low mass donors to
fill their Roche lobes.

{\em (iii)} It happens that in the low mass regime of donor mass where
the observational peak appears some of our synthetic systems are found
to be persistent XRBs. Below we give example of such evolutionary
scenario in the enthalpy model framework.

A binary after forming a BH ($8\msun$) enters a RLOF with a
$\sim1\msun$ donor at separation of $a=5\rsun$. The mass transfer is
driven by GR and MB for about $1-2\gyr$ at low rate ($\dot{M} \approx
10^{-10} \msy$) producing a transient behavior. In this time the
mass of a donor changes slowly (the peak in the companion mass
distribution $\sim1\msun$). With mass transfer the orbit continuously
decreases and once it is significantly reduced ($a\sim 2-3\rsun$) the
mass transfer rate increases and the system leaves the peak and moves
to the left in the mass distribution. Once mass of the donor drops
below $\sim 0.7-0.6\msun$ the mass transfer rate ($\dot{M} \gtrsim 5
\times 10^{-10} \msy$) exceeds the critical rate for disk instability and 
the binary becomes a persistent XRB. Within short time ($\sim
0.5\gyr$) the donor loses most of its mass and its evolution becomes
irrelevant.

\subsection{Angular momentum model}\label{sec:rnel}

\begin{figure}
    \centering
    \includegraphics[scale=0.6]{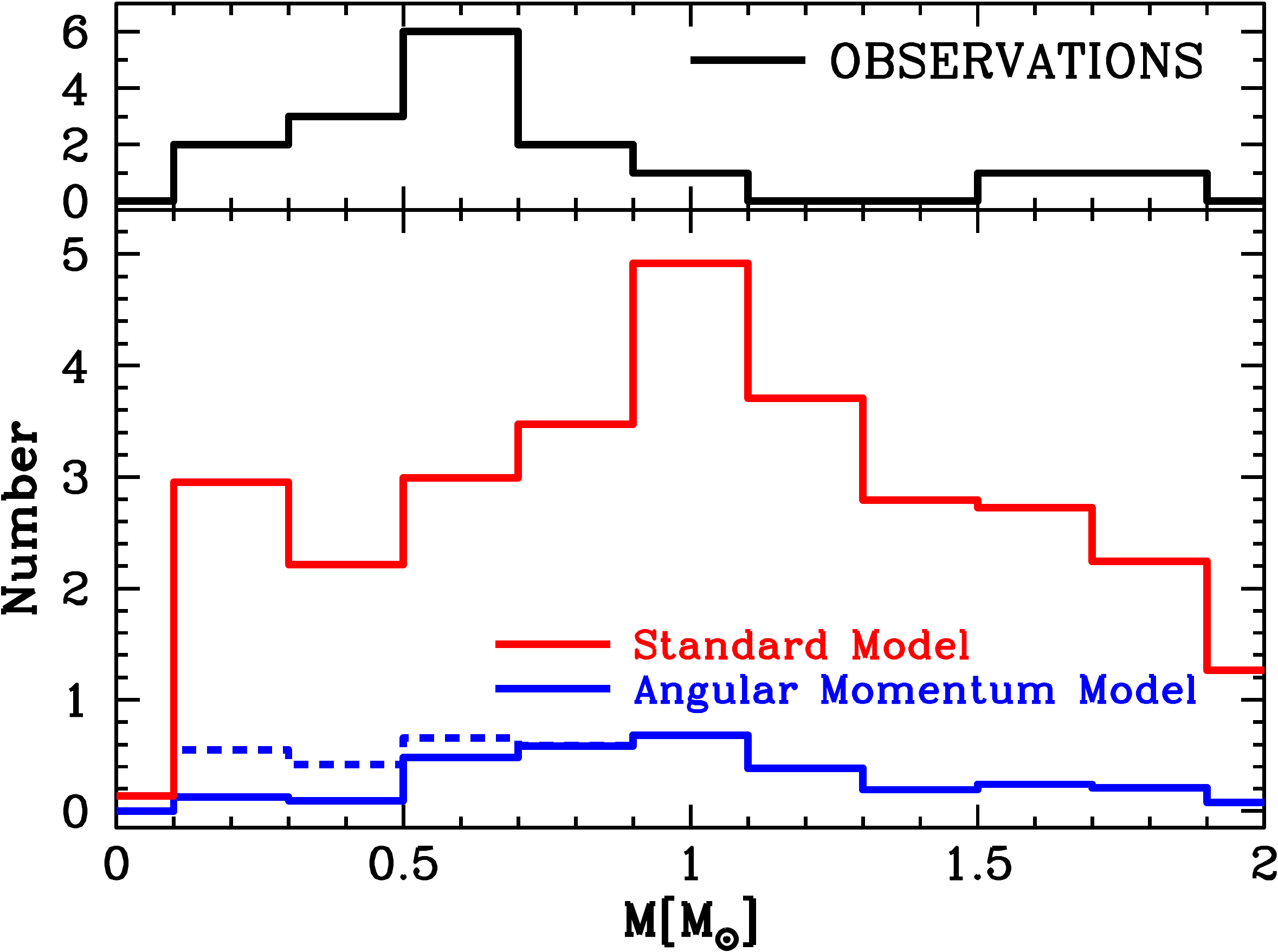}
    \caption{
The distribution of donor mass in BH systems for an evolutionary scenario
that employs angular momentum model of common envelope. Notation is the same
as in previous figures. In this model we obtain only about $3$ transient systems 
with low mass companions. Additionally, note that the predicted distribution
does not resemble the observations. 
    }
    \label{fig:nel}
\end{figure}

The companion mass distribution of Galactic BHXRT obtained with CE
prescription based on angular momentum
conservation \citep{Nelemans2000} differs significantly from the one
calculated with the standard model (see Figure \ref{fig:nel}). The
distribution is basically flat with most systems distributed in mass
range $0.5-1.3\msun$ with a tail extending to higher masses. There is
also a notable (by a factor of $\sim 10$) decrease in number of
predicted BHXRTs.

The initial evolution of binaries that enter BHXRT phase is vary
similar to the one in the standard model. The differences appear after
CE phase. Systems emerging from CE are much wider ($a\sim50\rsun$)
than in standard model ($a\sim5-10\rsun$). At such distances GR is not
effective in orbital decay. The combined action of MB (donors below
$1.25\msun$) and nuclear evolution are required to bring companions to
RLOF over a long period of time ($\sim5-10\gyr$). Relatively small
expansion on MS (factor of $\sim 2$) is not effective enough. Only
increased expansion on red giant branch (RGB) can initiate RLOF and
X-ray binary phase.  Only when the orbit is reduced to about half its
size by MB, a RGB donor starts RLOF.  The mass transfer rate is rather
constant ($\dot{M} \approx 10^{-8} \msy$) as a RGB star
properties depend primarily on its core mass ($\sim 0.35\msun$) and
not on its entire mass that decreases from $\sim 1-1.2\msun$ down to
its core mass. That explains rather flat companion mass distribution
over wide range of masses.

The very low BHXRT numbers in this model are the result of our
assumption on the outcome of CE with HG donors. Since this model
provides only moderate orbital contraction during CE the preference is
given to systems that enter the CE at the smallest orbital
separation. The massive stars that form BHs in our simulations evolve
from MS to HG and then directly to core helium burning (CHeB)
expanding at each phase.  We do not allow for survival of CE phase for
MS and HG donors due to potential lack of clear core-envelope
structure \citep{Belczynski2007}. Only binaries with wide orbits
that do not allow for RLOF before CHeB may survive CE phase and form
BHXRT. In this model only a small fraction of binaries that at the
very beginning of CHeB (the smallest allowed separations) initiate CE
can produce post-CE orbits small enough that BHXRT is formed within
Hubble time. In comparison, for the standard model evolution with more
efficient CE orbital contraction more binaries (over a wider range of
separations) can produce BHXRTs.

\subsection{Enhanced mass transfer model}\label{sec:rill}

\begin{figure}
    \centering
    \includegraphics[scale=0.6]{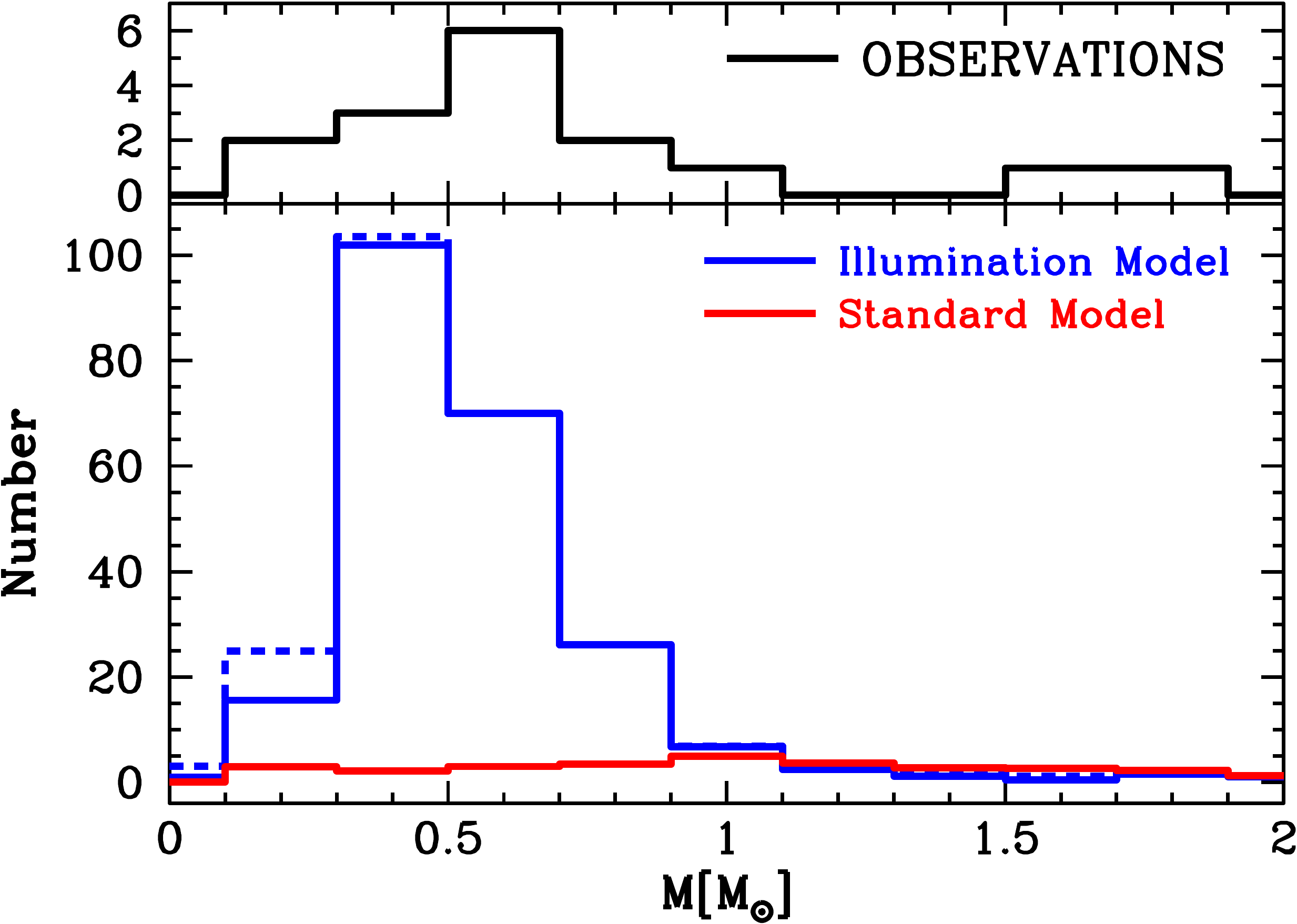}
    \caption{
The distribution of donor mass in BH systems for evolution with increased (by 
factor of $f_lambda=5$) mass transfer rate. The increase may be potentially 
caused by illumination of the companion by the accretion disk around BH. 
Note the large number of transient systems: $\sim 230$ and the fact that companion 
mass distribution peaks at $\sim0.4\msun$. Despite the fact that predicted donor
masses are rather close to the observed ones, this model is excluded based on the 
associated BH mass distribution. Majority of BHs in this model form 
via accretion induce collapse of heavy NSs and are found with mass just 
above $\sim3\msun$. This excludes this model as Galactic BHs are found with 
mass $5-15\msun$. See Section \ref{sec:rill} for more details. Notation is the same 
as in previous figures. 
    }
    \label{fig:ill}
\end{figure}

This model companion mass distribution shows the closest resemblance
to the observed distribution. We note a well defined peak at companion
mass $\sim 0.4\msun$ (see Figure \ref{fig:ill}). In this model we have increased
mass transfer rate as to shift the standard model peak of companion
mass from around $\sim1\msun$ toward lower, observed, values. It may
appear that we have quite successfully reproduced the observations
(peak at $\sim0.6\msun$).  However, this is not the case.

So far we have not directly commented on the associated BH mass
distribution of BHXRTs.  In all previously discussed models BH masses
were found in $5-15\msun$ range. This range is in agreement with the
existence of mass gap and consistent with masses of BHs in the
Galactic binaries \citep{Belczynski2011}. In this model we find that
majority of BHs have mass around $\sim3\msun$. This finding is
inconsistent with the observations. A potential observational bias
that may lead to an overestimate of BH mass was proposed
by \cite{Kreidberg2012}. However, it is hard to imagine that currently
known wide spectrum of BH masses would be shifted to a very narrow
range with peak at $\sim3\msun$.

Due to increased mass transfer the BHs in this model form primarily
from (more numerous) NSs via accretion induced collapse (AIC). The
typical evolution of such system is briefly described below.

At ZAMS a binary consists of $\sim20\msun$ and $\sim4\msun$
stars. Primary evolves quickly and initiates a CE (post-CE separation
$a=30\rsun$) and then forms a massive NS: $1.8\msun$. At
some point the mass transfer from a companion onto a NS
begins. At first the mass transfer rate ($\dot{M} \approx
10^{-5} \msy$) proceeds over critical Eddington rate
($dM_{edd}=1.7\times10^{-8}$) and most of the mass leaves the binary.
As donor mass decreases, the mass transfer slows down
($\dot{M} \approx 10^{-8} \msy$) and continues fully
conservatively for a prolonged time (about few $\gyr$).  During this
phase the accreting object reaches our adopted maximum NS mass
$M_{max,NS}=3.0\msun$.  At this point we assume that the NS
collapses to become a BH.  BH keeps accreting mass from the
companion to reach $\gtrsim3\msun$. After a compact object becomes a
BH, companion mass changes from about $0.7\msun$ to $0.35\msun$
producing the peak in the mass distribution.

In Figure \ref{fig:ill} we show results of calculation for the mass
transfer rate increase by factor of $f_{\rm ill}=5$. No qualitative
changes (in respect to BH mass distribution) are found for smaller
($f_{\rm ill}=2$) and larger ($f_{\rm ill}=10$) increase factors that
we have tested.

\subsection{Importance of maximum NS mass}\label{sec:rns2}

\begin{figure}
    \centering
    \includegraphics[scale=0.6]{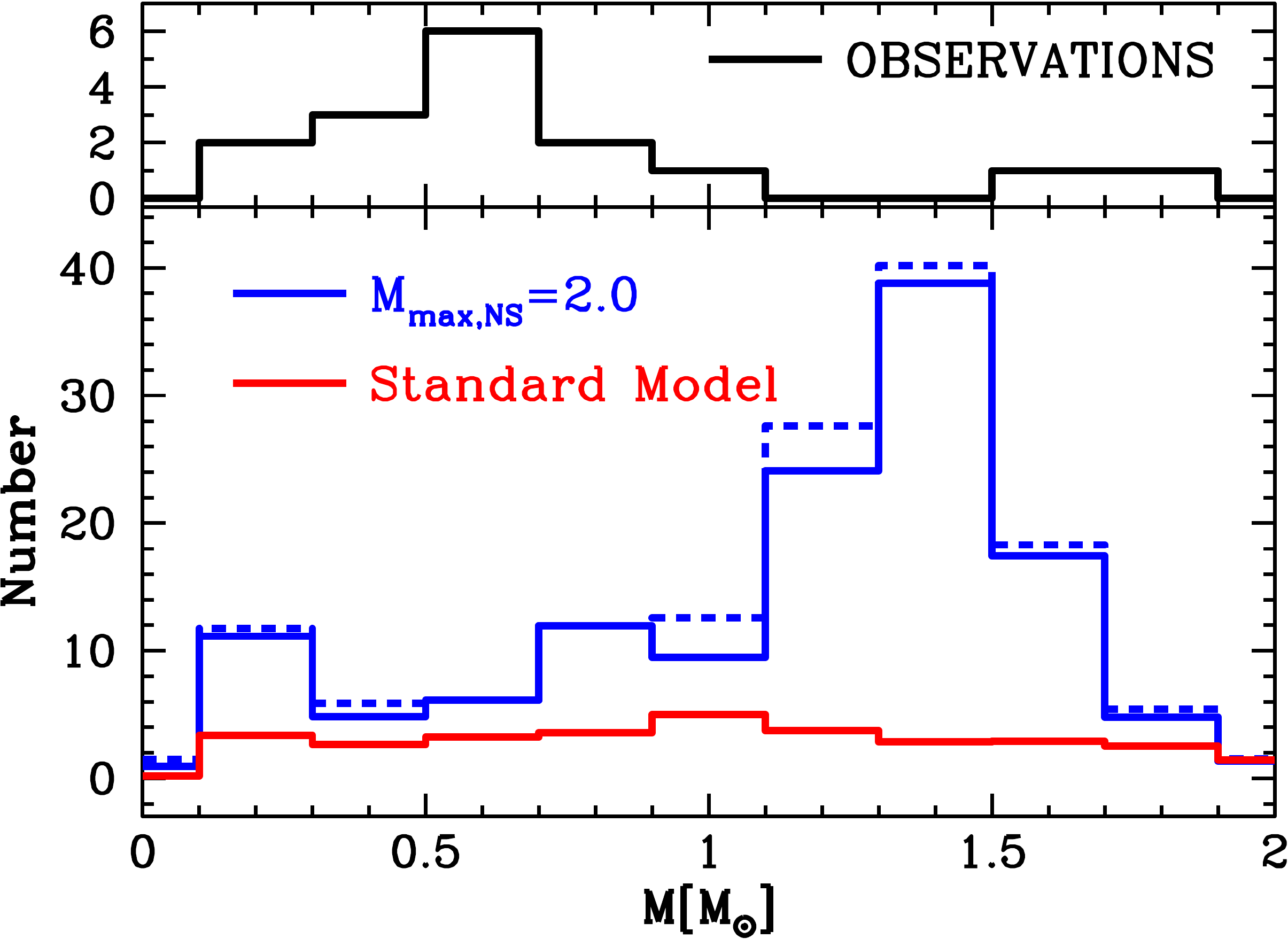}
    \caption{
The distribution of donor mass in BH systems for evolution with decreased
maximum NS mass: $M_{max,NS}=2$. In all the other models we have
employed a larger value $M_{max,NS}=3$. Note the increased number of BH
transients ($\sim130$) and shift of the distribution peak to $1.4\msun$ as
compared with the standard model. In this model we find majority of BHs 
with very low mass that have formed via accretion induced collapse. This and
the shape of companion mass distribution render this model as very unlikely. 
For more information see Section \ref{sec:rns2}. Notation is the same
as in previous figures.
    }
    \label{fig:ns2}
\end{figure}

The calculations presented above for different models employed maximum
NS mass of $M_{max,NS}=3\msun$ (see Section \ref{sec:startrack}). In
this model we lower this limit to $M_{max,NS}=2\msun$ and the
resulting companion mass distribution is presented in
Figure \ref{fig:ns2}. The distribution shows a very prominent peak at
$\sim1.4\msun$ with a minimum around the locus of observational
distribution maximum ($\sim0.5\msun$).

In this model we encounter the very similar situation as in the
enhanced mass transfer rate model. Although the mass transfer is not
increased, we have lowered the mass limit for BH formation. It
allows some NSs in accreting binary systems to increase
their mass just above $2\msun$, collapse to a BHs and become
BHXRTs. Since the typical companion mass is around $\sim1-3\msun$
there is just enough mass reservoir to build up BH mass to
$3-5\msun$. This disagrees with the observed BH mass distribution
($M_{\rm BH}\sim5-15\msun$) for the Galactic BHXRTs. We can exclude
this model based on the disagreement of the predicted BH and
companion mass distributions with observations. Below we give a brief
description of the typical evolution of the system in this model.

Initial progenitor binaries are qualitatively very different than in
the standard model.  Typical system consists of $\sim20\msun$ primary
and $\sim2-3\msun$ secondary. At $10$Myr since ZAMS, after CE and
supernova explosion a highly eccentric binary forms ($a=70\rsun$,
$e=0.9$) with a relatively heavy NS ($M_{\rm NS}\sim1.7\msun$). Tidal
forces are rather ineffective for stars with radiative envelopes
(e.g., \citet{Claret2007}) and it is not expected that such a binary
will circularize before the secondary star overflows its Roche lobe
(nuclear expansion on MS) at periastron passages. This happens at
about $100$Myr after the supernova explosion and we circularize the
binary at periastron distance ($a=7\rsun$, $e=0$). The mass ejected
from the secondary at periastron passages must put some drag on a
companion and lead to gradual circularization (however
see \citet{Sepinsky2010}).  The system starts its evolution through a
RLOF. First, we encounter a phase of thermal-timescale RLOF
($\dot{M} \approx 10^{-6} \msy$) and then the mass transfer
quickly slows down and allows for steady accretion onto NS at moderate
rates ($\dot{M} \approx 10^{-8}-10^{-10}\msy$). At about $1\gyr$
NS goes through AIC and forms a low mass BH ($M_{\rm
BH}=2\msun$). Companion mass is at that point reduced to $1-2\msun$
and system appears at BHXRT with moderate-to-low mass transfer rate
($\dot{M} \approx 10^{-10}\msy$) and on very close orbit
($a\sim4-5\rsun$). As mass transfer continues to remove mass from the
donor, MB turns on and this combined with increasing strength of GR
($a\sim3\rsun$) leads to further orbital decay. At the same time the
donor evolves beyond MS and its radius increases.  This leads to
unstable situation and most likely the merger of the two stars. We
stop calculations when donor radius exceeds its Roche lobe radius by
factor of $2$ at $\sim2 \gyr$ since ZAMS.

\subsection{Other factors}\label{sec:other}

Besides modifications to the input physics described above we have
tried to alter several assumptions and employed laws/formulas that are
important in modeling of Galactic BHXRTs. Below, we list the
additional tests that we have performed.

We have employed two other prescriptions for magnetic braking, with \citep{Andronov2003} 
and without \citep{Rappaport1983} saturation of magnetic dynamo. Along the similar 
lines we have narrowed down the appearance of convective envelopes for MS stars from 
$0.35-1.25\msun$ to $0.35-0.9\msun$, thus suppressing the action of MB for stars in the 
main peak of most predicted companion mass distributions. 
In several models we have exchanged our assumption on continuous star formation rate in
Galaxy with other prescriptions in order to assess the influence of BHXRT lifetime on the 
companion mass distribution. We have assumed that entire Galactic star formation was 
contained within $1\gyr$ burst and we have positioned the burst at four different times 
beginning at $10$, $9$, $2$, $1 \gyr$ ago. Then we have assumed that star formation
was linear function of time. In one model we have assumed that star formation starts 
$10\gyr$ ago at such value that it drops down to zero at present. In the other we have 
assumed that it starts $10\gyr$ ago at zero and at present increases to its maximal value. 
All models were normalized in such a fashion as to obtain total stellar mass of 
$3 \times 10^{10}\msun$ formed over entire $10 \gyr$ of the Galactic disk evolution.
None of the calculated companion mass distributions found in above simulations resembled 
the observed distribution of donor mass in Galactic BHXRTs.

\begin{figure}
    \centering
    \includegraphics[scale=0.6]{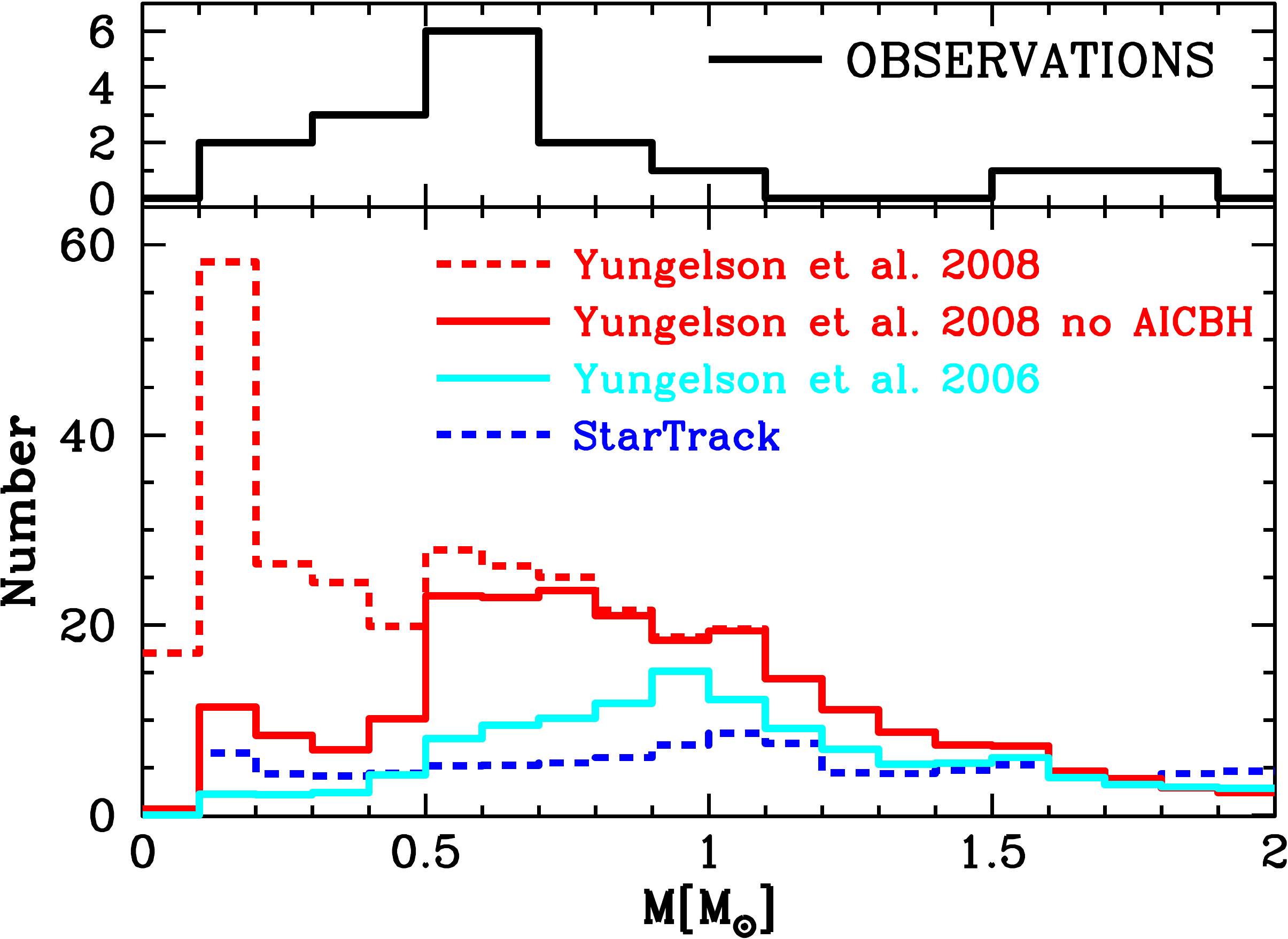}
    \caption{
The distribution of donor mass in BH transient systems obtained whit the \startrack 
model modified to resemble recent {\tt Seba} calculations. Model marked as 
\citet{Yungelson2006} have very high CE ejection efficiency and extended initial 
donor mass range for MB to operate. In model denoted as \citet{Yungelson2008}
we have additionally suppressed MB during ongoing RLOF. For this model we 
note the significant population of light BHs ($\sim 2-3\msun$) formed through 
accretion induced collapse of NSs. Our standard {\tt StarTrack} model is
shown for comparison.  See Section~\ref{sec:other} for details. 
}
    \label{fig:yun}
\end{figure}

We have changed parameters in \startrack to make it more similar to the {\tt SeBa} 
population synthesis code described in \citet{Portegies1996} with later updates in 
\citet{Portegies1998}, \citet{Nelemans2001} and \citet{Nelemans2004}. In particular, 
we have extended mass range for MS stars for which MB operates with to $0.3$--$1.6\msun$, 
for common envelope we have used energy balance with constant $\alpha \times \lambda=2$, 
primary initial mass range was narrowed down to $25$--$100\msun$, while secondary mass 
down to $0.08$--$100\msun$ and we have adopted slightly higher natal kicks for neutron 
stars (Maxwellian with $\sigma=300$ km s$^{-1}$). BH kicks are obtained by scaling down 
NS kicks. The rest of the parameters are the same as in our standard model calculation. 
This model is similar to the calculations of \citet{Yungelson2006} and we show the 
resulting donor mass distribution in Figure~\ref{fig:yun}. There is a slight improvement with the peak 
of distribution now at $\sim 0.9\msun$ and higher number of BH transients than found in 
our standard calculation. This is a direct result of adopting $40$-times higher CE 
efficiency ($\alpha \times \lambda=2$) as compared to our standard calculation ($\alpha
\times \lambda=0.05$).  

In next step, we employed the same modification to our standard model as described above 
and we additionally follow \citet{Yungelson2008} suggestion and (totally) suppress  
magnetic braking during the ongoing RLOF. The results are shown in Figure~\ref{fig:yun}.
The peak of donor mass distribution is now found at $\sim 0.5-0.6\msun$, very close to 
where the observed peak appears ($\sim 0.5-0.7\msun$). Since suppression of MB extends 
the lifetime of BH transients, we also find many more systems in this model as compared 
to our standard calculation. 

Systems that are found in the peak of donor mass distribution ($\sim 0.5-0.6\msun$) form 
from massive primaries and secondaries within initial mass range $\sim 0.8-1.25\msun$. 
After CE and BH formation, the orbital separation decreases due to combined action of MB 
and GR. Typically after several \gyr the secondary begins RLOF. We then suppress MB and 
the rate of angular momentum loss is significantly reduced. Companion slowly reduces its 
mass and with decreasing mass the mass transfer slows down. During the Hubble time, the 
star seldom falls below the $\sim0.5\msun$. Stars with masses lower than $0.8\msun$ on 
ZAMS rarely have enough orbital energy to eject the envelope of massive primary and 
survive CE phase. Stars heavier that $\sim1.6\msun$ are not subject to MB thus the 
secondary needs to finish CE very close to primary to be brought to RLOF by GR alone. As 
a result, the range of possible pre-CE separations is more restricted, which leads to 
lower formation probability of systems with donor mass over the MB threshold. 

There is an additional population present in this model (see dashed line in Figure~\ref{fig:yun} marking 
systems with typical donor mass $\sim 0.1\msun$). As it appears there is a significant 
population of systems with light compact objects $\sim 2-3\msun$ and very low-mass 
companions. These systems form at first heavy NS ($1.6-1.9\msun$) that later during RLOF 
from a relatively massive companion ($1.5-1.7\msun$) collapses via AIC to become a light 
BH. If MB was not suppressed during RLOF then NS would not have accreated enough mass to 
become a BH as the two stars would very quickly merge due to very effective angular 
momentum loss. Since transients does not seem to host light BHs then, within the framework of 
our model, this result indicates that MB is not suppressed during RLOF. The assumption on MB 
suppression is further disapproved by recent observations of XTE J1118+480 and A0620-00
and they very rapid orbital period decay indicating very strong MB for close RLOF BH 
transients \citep{Gonzalez2014}. Finally, in our opinion, this model involves unrealistically 
high CE efficiency ($\alpha=40$; if the physical estimates of binding energy $lambda=0.05$ is 
used).

\section{Selection effects on the X-ray binary population}\label{sec:selection}

Given that there are some disagreements between all of the theoretical
models here and the observed distribution of X-ray binaries, some
discussion of the selection effects for BH X-ray binaries is
necessary.  For objects to appear in the sample of 19 dynamically
confirmed stellar mass BHs, they must meet two key criteria:
(1) they must be detected in the X-rays and (2) they must have optical
or infrared counterparts bright enough and in regions of low enough
crowding that their mass functions can be measured.

The first criterion may be the source of the lack of objects with very
low mass donor stars.  Low mass donor stars will necessarily be main
sequence stars (or degenerate low mass objects).  Given the
period-mass relation for Roche lobe overflowing objects (see \citet{Knigge2011}
 for the most
up-to-date discussion of this relation), donor stars of 0.2 $M_\odot$
should overflow their Roche lobe in systems with orbital period of
about 2 hours.  

A correlation between the peak X-ray luminosity of an X-ray transient
and the orbital period of the transient has been well established in
recent years (\citet{Shahbaz1998};
\citet{Portegies2004}; \citet{Wu2010}).  At the shortest orbital periods, the peak accretion rates
will be well below the threshold of a few percent of the Eddington
rate where BH X-ray binaries enter the high soft state
\citep{Maccarone2003}.  As a result, the systems are likely
to be in advection dominated states, with two consequences -- first
that the accretion will be radiatively inefficient, reducing the
bolometric luminosity, and second that the X-ray spectrum will peak at
100-200 keV, rather than at a few keV, meaning that the most sensitive
all-sky monitors, which have traditionally operated below 15 keV, will
not catch the peak of the X-ray spectrum.  \citet{Maccarone2013}
noted that these objects might manifest themselves as ``very faint
X-ray transients'', while \citet{Knevitt2013} have discussed how
these objects might be absent in existing catalogs of X-ray binaries
for similar reasons.  It is worth noting that in recent years, a few
objects have been discovered that are strong (but not yet dynamically
confirmed) BH candidates, and which have orbital periods less
than 4 hours -- Swift J1753.5-1027 \citep{Zurita2008}, and MAXI J1659-152 \citep{Kuulkers2013}, and
these were both discovered by All-Sky Monitors much more sensitive
than those which existed before Swift was launched.  It is important
to note, also, that selection effects on the basis of orbital period
are much more likely to be important than selection effects on the
basis of BH mass -- see \citet{Ozel2010} for a discussion
of why the latter is, at most, a minor issue.

The second criterion is also a potentially serious issue, although how
it may manifest itself is not entirely clear.  This paper presents a
study of 19 objects which are dynamically confirmed BH
candidates.  A slightly larger number of objects show strong albeit
indirect evidence for being BHs \citep{Remillard2006}.  These 
lines of evidence can include: systems showing an
ultrasoft spectral component \citep{White1984};
showing a lack of Type I X-ray bursts despite having been well-studied
\citep{Remillard2006}; and strong radio emission
relative to the X-ray emission level (e.g. \citet{Strader2012}
; \citet{Chomiuk2013}.  In one case,
that of 4U~1957+11, a system shows a range of indirect evidence for
being a BH, but has remained as a persistently bright source
since the Uhuru era, making it impossible for its donor star's light
to be seen, and hence for its mass function to be estimated
(e.g. \citet{Nowak2008}; \citet{Russell2011}). 
In the two globular clusters, the crowding of the stars
makes it difficult to measure the donor star's spectrum.  In nearly
all of the other cases, the systems are very close to the Galactic
Plane, so that foreground extinction makes measurement of the donor
stars' spectra impractical with current instrumentation.  The effects
of this criterion on the orbital period distribution of dynamically
confirmed BH X-ray binaries have not yet been studied well
enough to determine how it will manifest itself.  It seems likely,
though, that if there is a strong correlation between orbital period
and natal kick velocity, that this effect may be important -- being
kicked well out of the Galactic Plane makes optical follow-up much
easier.  An alternative to understanding the selection effects for
outbursting sources will be to develop and understand a sample of
quiescent BH X-ray binaries, one of the key goals of the
Chandra Galactic Bulge Survey \citep{Jonker2011}.

\section{Conclusions}\label{sec:conclusions}

We have reexamined the issue of donor mass in the Galactic BH
X-ray transient binaries. Since the formation scenarios involve CE
phase initiated by a massive BH progenitor, it is naturally expected
that companion mass should not be too small as to avoid the CE
merger. However, the donors that are found in Galactic BHXRTs have
very low mass $\sim0.6\msun$. Early studies have shown that stars with
mass above $2\msun$ are the most likely companions for Galactic
BHs. With the updated population synthesis code we have shown that
stars with mass $1\msun$ are most likely companions. Despite the
factor of $\sim2$ improvement the predictions are still in tension
with available observations.

\begin{figure}
    \centering
    \includegraphics[scale=0.6]{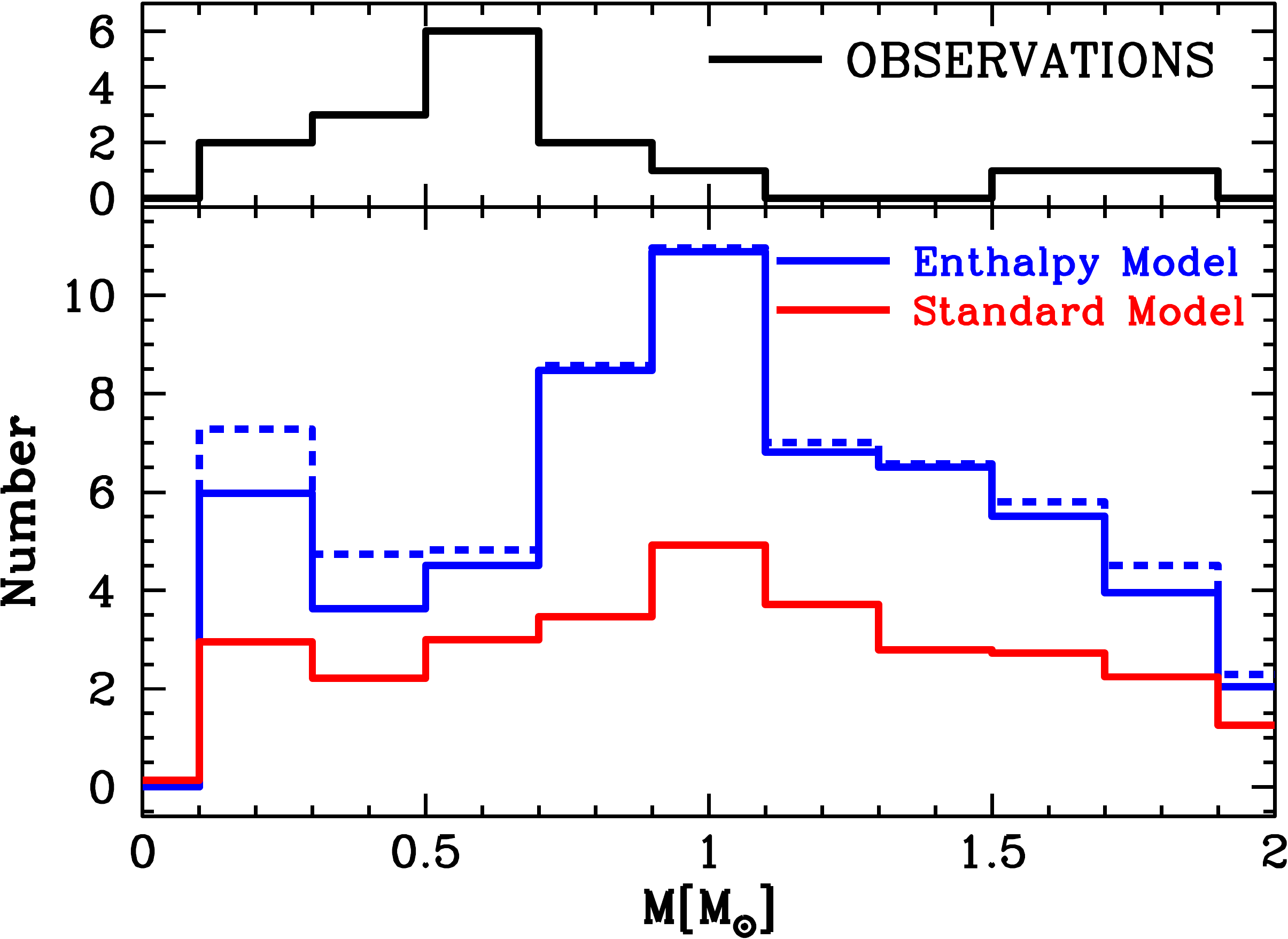}
    \caption{
The donor mass distribution for evolution that incorporates the enthalpy common 
envelop model with much easier envelope ejection that assumed in the standard model.  
We find  about $60$ BH transients (solid line) with low mass companions in 
this model. The persistent systems are also marked as in Figure \ref{fig:std}. 
The observations and standard model results are shown for comparison. 
Note that this model, despite the expectations, produces donors with a typical mass 
same as in standard model: $\sim1\msun$. For explanation see Section \ref{sec:rivn}.
    }
    \label{fig:ivn}
\end{figure}

We have implemented several alternatives and modifications in our
evolutionary calculations to test whether it is possible to bring
predictions closer to the observations. We have failed to reproduced
the observed distribution of companion mass. The problem seems to be
deeper than previously believed. Our results show that, in spite of the
fact that common envelope phase seems necessary for decreasing the separation
and, therefore, the formation of low mass X-ray binaries, it is not the 
crucial factor. This is demonstrated by comparison of two models;
one with the standard CE ejection efficiency and one with the
significantly increased efficiency (see Figure~\ref{fig:ivn}). There is 
no clear improvement in the position of the donor mass peak. Actually, both
models show minimum at the locus of the observed companion mass peak
($\sim0.6\msun$). The minimum found in almost all our models is a
consequence of a donor size. For these very low masses the donor is
small. To fill its Roche lobe a donor needs to be in a very close
binary. The very small separation means strong binary angular momentum
loss due to emission of gravitational radiation. It also sets the
donor fast rotation (synchronization assumed during RLOF) leading to
efficient magnetic braking. Both processes increase mass transfer rate
and therefore low mass donors lose mass faster than higher mass
stars. It means that the low mass stars from the observational peak
($\sim0.6\msun$) are less likely to appear in the BH binary population
than the higher mass stars, like the ones found in the predicted peak
($\sim1\msun$). Additionally, some systems with these low mass donors
disappear altogether as the high mass transfer rate stops the
transient behavior.

There are two general types of solution to this persisting problem. 
Either the observed distribution of companion mass is heavily biased 
or the details of mass transfer for low mass stars are not understood.

There are several factors that may potentially bias the observed
donor mass spectrum. There exists a potential bias in measurements of
binary inclination in BH transients that may affect component
mass estimates. However, the potentially underestimated inclinations
\citep{Kreidberg2012} would lead to a decrease in donor mass.
There are clearly factors which can make systems with low mass donor
stars harder to detect. There may also potentially exist a trend that
makes systems with higher mass companions less visible. For example,
for more massive donors with lower mass transfer rates it may take
longer for the disk to refill after the outburst. It would produce
longer duty cycle and thus decrease the probability of discovery as BH
candidates are generally identified by outbursts. Finally, it is not
well understood what effects natal kicks have on the difference
between the intrinsic and observed donor mass distributions.
Given that about half of the BH X-ray binaries are too heavily
extincted to have their system parameters measured, this is clearly
something that could potentially be important.
 
On the theoretical side the evolution through mass transfer for these
extreme mass ratio systems with very low mass companions may not be
fully understood. As the side effect of our main work, we
found that the very evident problem is the lack of generally
accepted magnetic braking model. More than that, it seems like all
proposed models fail to explain (MB too weak) the recent observations 
of rapid orbital decay of XTE J1118+480 \citep{Gonzalez2013}. Additionally, 
it is not clear whether the transient behavior can be estimated based only on
the mass transfer rate (and how uncertain such estimates are) as the
disk instability theory is far from being fully understood.

Besides various models for MB, with and without dynamo saturation, we applied 
significantly increased RLOF rates (by factor of $\sim 10$) as motivated by 
potential irradiation of donor by the accreting BH/disk. However, it may be also 
a proxy for increased MB operating for all our RLOF systems. We have also employed 
the idea put forward by \citet{Yungelson2006} and \citet{Yungelson2008} and 
we completely turned off MB for all RLOF systems. In both cases it was found that 
donor mass in our predicted population of Galactic BH transients very closely 
resemble observations. However, the BH mass spectrum is totally different (mostly 
light $\sim 2-3\msun$ BHs) than observed. It may very well indicate that more 
complex MB modification is required. Modification that not only alters the MB 
strength but that also most likely changes with properties of donor stars.
It may be potentially possible to reverse engineer MB from donor mass observations, 
provided that the observations are not heavily biased. Any such attempt
should take into account the available constraints from orbital decay measurements 
indicating that for very low mass RLOF donors ($\sim 0.2-0.4 \msun$) the MB
is much stronger than any available model can predict.

\acknowledgements

We would like to thank Tassos Fragos, Natasha Ivanova and Philipp
Podsiadlowski for useful suggestions and discussions. 
We would like to thank the Copernicus Astronomical Center in Warsaw,
Poland, and the University Of Texas, Brownsville, TX, and the Polish
PL-Grid project for their courtesy, enabling us to use their
computational resources. We acknowledge the Texas Advanced Computing Center at the University of Texas at Austin for providing HPC resources that have contributed to the results presented in this paper. This study was partially supported by the
Polish NCN grant N203 404939, Polish FNP professorial subsidy
"Master2013" and by Polish NCN grant SONATA BIS 2. KB also acknowledges 
NASA Grant Number NNX09AV06A and NSF Grant Number HRD 1242090 awarded to 
the Center for Gravitational Wave Astronomy, UTB.

\bibliographystyle{apj}
\bibliography{ms}

\newpage
\begin{center}
\begin{deluxetable}{rlccccc}
    \tablewidth{500pt}
    \tablecaption{Properties of Galactic Black Hole X-ray Binaries}
    \tablehead{No& Name  & $\mathrm{M}_\mathrm{comp}[\msun]\tablenotemark{b}$ 
    & Spec. type & $\mathrm{M}_\mathrm{BH}[\msun]$ & $\mathrm{P}_\mathrm{orb}$ & citations\tablenotemark{e}}
    \startdata
1 &    XTE J1118+480     & $0.22\pm0.07$         & K7/M1V & $6.9\div8.2$       & $4.08$          & [1,17,17,38] \\
2 &    XTE J1550-564     & $0.3\pm0.07$          & K2/4IV & $10.5\pm1.0$       & $37$            & [2,2,28,2]\\
3 &    GS 2000+25        & $0.16\div0.47 (0.315)$& K3/6V  & $\sim6.55$         & $8.26$          & [3,18,3,40]\\
4 &    GRO J0422+32      & $\sim0.45$            & M0/4V  & $\sim10.4$         & $5.09$          & [4,19,4,4]\\
5 &    GRS 1009-45       & $\sim0.5$             & G5/K0V & $\sim8.5$          & $6.86\pm0.12$   & [5,20,5,41]\\
6 &    GRS 1716-249      & $\sim1.6$             & K-M    & $\gtrsim4.9$       & $14.7$          & [37,37,37,37]\\ 
7 &    GX339-4           & $0.3\div1.1 (0.54)$   & KIV    & $>7$               & $42$            & [6,21,6,21]\\
8 &    H1705-25          & $0.15\div1.0$         & K3/M0V & $4.9\div7.9$       & $12.55$         & [7,22,29,29]\\
9 &    A0620-00          & $0.68\pm0.18$         & K2/7V  & $6.6\pm0.25$       & $7.75$          & [8,23,30,42]\\
10&    XTEJ1650-50(0)    & $0.7$                 & K4V    & $\sim5.1$          & $7.63$          & [\tablenotemark{c},24,31,24]\\  
11&    XTEJ1859+226      & $0.7$                 & K5V    & $7.7\pm1.3$        & $6.58\pm0.05$   & [\tablenotemark{c},25,32,25]\\
12&    GS2023+338        & $0.5\div1.0 (0.7)$    & K0/3IV & $12\pm2$           & $156$           & [9,26,33,43]\\
13&    GRS 1124-68       & $0.3\div2.5 (0.8)$    & K5V    & $6.95\pm0.6$       & $10.392$        & [10,10,34,44]\\
14&    GRS1915+105       & $0.8\pm0.5$           & K1/5III& $12.9\pm2.4$       & $811.2\pm2.4$   & [11,27,35,35]\\
15&    GS 1354-64        & $1.03$                & G5IV   & $7.6\pm0.7$        & $61.07$         & [\tablenotemark{c},12,12,12]\\
16&    GROJ1655-40       & $1.75\pm0.25$         & F3/G0IV& $5.31\pm0.07$      & $62.909\pm0.003$& [\tablenotemark{d},13,36,45]\\
 & & & & & & \\
17&    4U1543-47         & $2.3\div2.6 (2.45)$   & A2V    & $2.7\div7.5$       & $26.8$          & [14,14,14,46]\\ 
18&    XTEJ1819-254      & $5.49\div8.14 (6.81)$ & B9III  & $8.73 \div 11.70$  & $67.62$         & [15,15,38,15]\\
19&    CygX-1 \emph{persistent} \tablenotemark{a}& $19.2\pm1.9$ & OI & $14.8\pm0.1$ & $134.4$    & [16,16,16,16]\\
    \enddata
    \label{tab:bhxrbs}
    \tablenotetext{a}{Cyg X-1 is the only persistent Galactic source, other systems are transients}
    \tablenotetext{b}{When only a range of companion masses is available the value used to built the 
      observational distribution (e.g., Figure~\ref{fig:std}) is given in parenthesis}
    \tablenotetext{c}{Derived from the spectral type}
    \tablenotetext{d}{Derived from $M_{BH}=5.31\pm0.07$\citep{Motta2014} and
    $q=0.329\pm0.047$\citep{Gonzalez2008}} 
    \tablenotetext{e}{Citations are organized as follows: [$\mathrm{M}_\mathrm{comp}$,Spectral type,$\mathrm{M}_\mathrm{BH}$,$\mathrm{P}_\mathrm{orb}$]. References: 
[1]\citet{Gonzalez2012}; 
[2]\citet{Orosz2011b};   
[3]\citet{Ioannou2004} ; 
[4]\citet{Reynolds2007}; 
[5]\citet{Macias2011}  ; 
[6]\citet{Munoz2008}   ; 
[7]\citet{Martin1995}  ; 
[8]\citet{Gelino2001}  ; 
[9]\citet{Casares1994};  
[10]\citet{Shahbaz1997};  
[11]\citet{Harlaftis2004}
[12]\citet{Casares2009} ; 
[13]\citet{Beer2002}    ; 
[14]\citet{Orosz1998}   ; 
[15]\citet{Orosz2001}   ; 
[16]\citet{Orosz2011a}  ; 
[17]\citet{Khargharia2013};
[18]\citet{Harlaftis1996};
[19]\citet{Gelino2003};
[20]\citet{DellaValle1997};
[21]\citet{Hynes2003};
[22]\citet{Filippenko1997};
[23]\citet{Froning2011};
[24]\citet{Orosz2004}   ; 
[25]\citet{Corral-Santana2011};
[26]\citet{Khargharia2010};
[27]\citet{Greiner2001};
[28]\citet{Li2013}      ; 
[29]\citet{Harlaftis1997};    
[30]\citet{Cantrell2010} ;    
[31]\citet{Sluny2008}    ;    
[32]\cite{Shaposhnikov2009};  
[33]\citet{Shahbaz1994}    ;  
[34]\citet{Gelino2001b}    ;  
[35]\citet{Hurley2013}   ; 
[36]\citet{Motta2014}   ;    
[37]\citet{Masetti1996}  ;      
[38]\citet{Martin2008};
[39]\citet{Torres2004};
[40]\citet{Chevalier1990};
[41]\citet{Shahbaz1996};
[42]\citet{Johannsen2009};
[43]\citet{Casares1992};
[44]\citet{Orosz1996};
[45]\citet{Gonzalez2008};
[46]\citet{Orosz2002};
}
\end{deluxetable}
\end{center}

\newpage

\newpage

\newpage

\newpage

\newpage

\newpage

\end{document}